\documentclass[showpacs,preprint,11pt,nofootinbib,citeautoscript,eqsecnum]{revtex4}
 \usepackage{amsmath,amsfonts,amssymb}
\usepackage{graphics}
\usepackage{color}
\usepackage{epsfig}
\usepackage{pdfpages}
\usepackage[font=small,labelfont=bf]{caption}
\definecolor{darkblue}{rgb}{0.0,0.0,0.8}
\definecolor{darkred}{rgb}{1,0.5,0.7}
\definecolor{Brown}{cmyk}{0, 0.8, 1, 0.6}
\usepackage{setspace}
\usepackage{hyperref}
\hypersetup{colorlinks=true, citecolor=darkblue, linkcolor=Brown,  urlcolor=darkblue}
\newcommand{\nc}{\newcommand}
\nc{\ba}{\begin{eqnarray}}
\nc{\ea}{\end{eqnarray}}
\newcommand\be{\begin{equation}}
\newcommand\ee{\end{equation}}

\nc{\x}{{\bf{x}}}

\input{epsf.sty}

\begin{document}

 \title{Graviton Propagation in an Asymmetric Warped Background: Lorentz Violation and the Null Energy Condition}

\author{Shahrokh Parvizi and Hossein Rezaee}
\affiliation{Department of Physics, School of Sciences, Tarbiat Modares University, P.O. Box 14155-4838, Tehran, Iran,\\ parvizi@modares.ac.ir,  hossein.rezaee@modares.ac.ir}

 \date{\today}

\begin{abstract}
The graviton propagation in an asymmetric background is studied. The background is a configuration in the six-dimensional Salam-Sezgin model, in which a 3-form H-field turned on [{\em JHEP} {\bf 0910} (2009) 086]. The compact dimensions form a cylindrical space with branes as boundaries. The background gets asymmetry due to the H-field and violates the Lorentz symmetry. We derive the graviton equation in this background and show that it gets massless mode traveling with superluminal speed. A tower of K-K modes exists with a mass gap. On the other hand, it is known that breaking the Lorentz symmetry on an asymmetric background is constrained by the null energy condition. This no-go theorem doesn't work well in six-dimensional space-times and by this model we provide a counterexample for which the null energy condition is satisfied while the Lorentz symmetry is gravitationally violated.
\end{abstract}

\maketitle
\tableofcontents

\section{Introduction}\label{introduction}
Higher dimensional theories have been at the center of interest in recent decades. They emerge as necessary ingredients of string theory and when utilized with branes and warped compactification, provide great phenomenological implications such as hierarchy problem in field theories \cite{ArkaniHamed:1998rs,Randall:1999ee,Randall:1999vf} or cosmological constant problem in gravitational theories \cite{ArkaniHamed:2000eg,Kachru:2000hf}. Embedding our 4-dimensional world as a brane in a higher dimensional spacetime brings us more chance to capture higher dimensions in accessible energy scales in high energy accelerators such as LHC.

For a long time, the Salam-Sezgin supergravity in 6-dimension \cite{Nishino:1984gk,Salam:1984cj,RandjbarDaemi:1985wc,Nishino:1986dc} has attracted attentions as a relatively simple model to study the warp compactifications \cite{Carroll:2003db,Navarro:2003vw,Navarro:2003bf,Aghababaie:2003wz,Aghababaie:2003ar,Gibbons:2003di,Burgess:2004dh}. It has the advantage that can be derived from string theory and has a consistent bosonic truncation \cite{Cvetic:2003xr}. The bosonic part of the model contains graviton, dilaton, a 2-form $F$, and a 3-form $H$ field. In most of brane-world scenarios, based on the Salam-Sezgin model, $H$ field was set to zero and a 4-dimensional Lorentz symmetric compactification was presented \cite{Carroll:2003db,Navarro:2003vw,Navarro:2003bf,
Aghababaie:2003wz,Aghababaie:2003ar,Gibbons:2003di,Burgess:2004dh,Lee:2005az, Copeland:2007ur,Tolley:2006ht,Burgess:2006ds,Lee:2008pz,Parameswaran:2006db,Parameswaran:2007cb}. The perturbation around this symmetric background and modification to the Newtonian gravity was studied in \cite{Parameswaran:2009bt,Salvio:2009mp}.

Including $H$ field was firstly done in \cite{Afshar:2009ps} where a static model obtained and searching for a dynamical metric was followed in \cite{Minamitsuji:2010fp,Minamitsuji:2010kb}. In \cite{Afshar:2009ps} an axially symmetric internal space was introduced, where the radial direction was cut by two 4-branes which wrapped over the azimuthal circle. Smeared 3-branes and zero-branes were also introduced to satisfy Israel junction conditions.

Since $H$ is a 3-form, turning it on presumably violates the Lorentz symmetry. Indeed in the presence of the $H$ field, an asymmetry shows up in the metric as the warp factors for time and space are different. This asymmetric warping had been studied before in different models, sometimes known as time warp \cite{Kaelbermann:1998hu,Visser:1985qm,Chung:1999xg,Kraus:1999it, Youm:2001sw,Kiritsis:1999tx,Alexander:1999cb,Bowcock:2000cq,Csaki:2000dm,Cline:2001yt,Cline:2003xy,Dubovsky:2001fj,Deffayet:2001aw,Frey:2003jq,Ganor:2006ub,Koroteev:2007yp,Gubser:2008gr}.
However, it was shown that the warp factor can be made to be symmetric at the physical brane which restores the Lorentz symmetry on the brane. This is interesting for standard model fields which are confined on the brane, but doesn't save the Lorentz invariance for gravity modes which inherently propagate in all directions including off the brane.
The Lorentz violation is claimed to be
one of the most efficient way to explore new physics and important to those who are curious in the relation of gravitational and quantum phenomena \cite{Kiritsis:2012ta}.

On the other hand, the importance of the model in \cite{Afshar:2009ps} is bypassing a no-go theorem originated from the null energy condition \cite{Cline:2001yt,Gubser:2008gr,Gubser:2011mp}. The no-go theorem states that the internal space for any asymmetric warp compactification in $D\neq 6$, indeed can not be compact, unless the null energy condition violated. So any Lorentz violation scenario based on higher dimensional gravity is restricted by the no-go theorem \cite{Gubser:2011mp}. The silence of the no-go theorem in $D=6$ dimension makes the model \cite{Afshar:2009ps} a candidate for the gravitational Lorentz violation in higher dimensions without violating the null energy condition.

In this article, we follow $H$ field model in \cite{Afshar:2009ps}, consider the spatial tensor perturbation of metric and derive gravitational wave equation. The equation is accompanied by boundary conditions at branes. Since it is too complicated to be solved analytically, we perform numerical analysis to find a solution. Results involve the graviton spectrum including a massless mode with a mass gap for higher modes. Positive definiteness of the spectrum indicates the stability of the model as long as tensorial perturbation is concerned. Phenomenologically, finding a massless state with a mass gap is interesting and shows that the effective four dimensional gravity can be obtained in this model. As expected, the graviton propagation generates an energy-momentum dispersion relation which violates the Lorentz symmetry. Our numerical results show that the phase velocity as $c=E/P$ in some range of energy exceeds the limit 1. This says that while the electromagnetic wave speed is already 1 on the brane-world, the gravitational wave speed limit is over 1 due to the asymmetric warp factor. This is an explicit example of gravitational Lorentz violation while the null energy condition is satisfied.

\section{The Set-Up}\label{setup}

\noindent In this section we give a brief introduction to the model in \cite{Afshar:2009ps}. Let us begin by the bosonic part of Salam-Sezgin Lagrangian as
\begin{equation} \label{setup-eq1}
      \frac{\mathcal{L}}{\sqrt{-g}}=\frac{1}{2\kappa^2} (-\mathcal{R}-
      \partial_M\phi\partial^M\phi)-\frac{1}{4}e^{-\phi} F_{MN}F^{MN}
      -\frac{1}{6} e^{-2\phi} H_{MNP}H^{MNP}-\frac{2g^2}{\kappa^2} e^{\phi}
\end{equation}
where $\phi$, $F$ and $H$ are respectively dilaton, 2 and 3-form fields.
The constant $g$ may also be recognized as the cosmological constant.
Equation of motion governing each field is obtained as follows,
\begin{subequations} \label{setup-eq2}
\begin{align}
      &-\mathcal{R}_{MN}=\partial_M \phi \partial_N \phi+\frac{g^2}{\kappa^2}
       e^\phi G_{MN}+\frac{1}{2} \kappa^2 e^{-2\phi} (H_{MN}^2-\frac{1}{6} H^2 G_{MN} ) \nonumber\\
      &\qquad \qquad \quad +\kappa^2 e^{-\phi} (F_{MN}^2-\frac{1}{8}F^2G_{MN}), \label{setup-eq2a} \\
      &\Box \phi+\frac{\kappa^2}{6} e^{-2\phi} H_{MNP} H^{MNP} +\frac{\kappa^2}{4}
      e^{-\phi} F_{MN} F^{MN} -\frac{2g^2}{\kappa^2} e^\phi =0,  \label{setup-eq2b} \\
     &D_M (e^{-2\phi} H^{MNP} )=0,  \label{setup-eq2c}  \\
     &D_M(e^{-\phi} F^{MN} )+e^{-2\phi} H^{MNP} F_{MP}=0.  \label{setup-eq2d}
\end{align}
\end{subequations}
We also take the space-time described by
\begin{align} \label{setup-eqn3}
      &ds^2 = -e^{2w(\eta)} dt^2+e^{2a(\eta)} \delta_{ij}dx^i dx^j+e^{2v(\eta)} d\eta^2
      +e^{2b(\eta)} \theta'^2,\nonumber \\
      &\qquad F = 0, \qquad e^{\phi} =e^{\phi(\eta)},
      \qquad H = h'(\eta) dt\wedge d\theta' \wedge d\eta.
\end{align}
For later convenience we take $z=\eta/l_z$ and $\theta = \theta'/l_{\theta}$.
Now $(z, \theta)$ are dimensionless cylindrical coordinates and $(l_z, l_\theta)$ stand for compactification radii of extra dimensions. 
Inserting the metric ansatz into field equations \eqref{setup-eq2}, a natural gauge condition for fixing parameter $z$ seems to be $w'+3a'-v'+b'=0$ that leads to following solution:
\begin{align}
      & h^{\prime}(z)=\pm q e^{2x}                                      \nonumber \\
      & w(z) = \frac{y + x}{4} + (2\lambda_3 + \lambda_4)(|z| + z_3)    \nonumber \\
      & a(z) = \frac{y-x}{4} - \frac{\lambda_3}{3}(|z|+z_3)             \nonumber  \\
      & v(z)=\frac{5y-x}{4} + \lambda_3(|z|+z_3)                        \nonumber \\
      & b(z)=\frac{y+x}{4} - \lambda_4 (|z|+z_3)                        \nonumber \\
      & \phi(z) = \frac{y-x}{4} - 2\lambda_3 (|z|+z_3)                  \label{setup-eq4}
\end{align}
where $q$, $\lambda_3$, $\lambda_4$ and $z_3$ are some constants, and auxiliary functions $x$ and $y$ satisfy
\begin{align}
      & {x^{\prime}}^2- 2 \tilde{q}^2 e^{2x} = \lambda^2, \nonumber \\
      & {y^{\prime}}^2 + 4 \tilde{g}^2 e^{2y} = 1,  \label{setup-eq5}
\end{align}
with $\lambda$ being another constant, $\tilde{q} = \kappa q/l_z$ and $\tilde{g}=gl_z/\kappa$ are now dimensionless. The absolute value of extra dimension originates from the fact that to avoid any singularity, one needs to cut the geometry, say between 0 and $L$, then double it to find a periodic solution between $-L$ and $L$. Finally we have a compactified space in $(0, L)$ interval with a $\mathbb{Z}_2$ symmetry and absolute values appear at boundaries. The topology of the internal space would be a cylinder with two boundaries.

Before going on the boundary conditions and introducing branes, let us make some comments on the solution and parameters involved. The general form of $e^{-x}$ from the solution of first equation is one of
$\textit{sinh}$/$\textit{sin}$ or $\textit{linear}$ functions corresponding to
the value of $\lambda^2$ to be $\textit{positive}$/$\textit{negative}$
or $\textit{zero}$, respectively. Here we take the positive sign and, without loss of generality, assume that $\lambda$ is positive as well.
Equations (\ref{setup-eq5}) then read
\begin{align} \label{setup-eq6}
      & e^{-x} = \frac{\sqrt{2} \tilde{q}}{\lambda} \left|\sinh(\lambda(|z| + z_1))\right| \nonumber \\
      & e^{-y} = 2 \tilde{g} \cosh(|z|-z_2),
\end{align}
where $z_i$'s are integration constants. Notice that solutions to \eqref{setup-eq5}  in the limit $\tilde{q}$ and $\tilde{g} \to 0$ are $x = \pm \lambda z+c_1$ and $y = \pm z+c_2$. However the hyperbolic functions
in Eqs.~\eqref{setup-eq6} could not essentially reduce to these limiting solutions, unless the constants $z_i$'s are  chosen properly.
This can be done by rewriting, for example, $e^{-x}$ in \eqref{setup-eq6} as
\begin{equation}
      e^{-x} = \frac{\tilde{q}}{\sqrt{2}|\lambda|} \left| e^{\lambda(|z| + z_1)}-e^{-\lambda(|z| + z_1)} \right|
      = \left(\frac{\tilde{q} e^{\lambda z_1}}{\sqrt{2}\lambda} \right)
      e^{\lambda |z|}\left| 1-e^{- 2\lambda(|z| + z_1)}\right|, \nonumber
\end{equation}
and then taking logarithm of both sides and let $\frac{\tilde{q}}{\sqrt{2} \lambda} e^{\lambda z_1}=1$, the limiting solution $x = \pm \lambda z+c_1$ can be achieved as $\tilde{q}\to 0$. In the same way, $\tilde{g} e^{- z_2}=1$. The solutions to Eqs.~\eqref{setup-eq5} are therefore:
\begin{align} \label{setup-eq7}
& x = -\lambda |z| - \ln  \left| 1 - \bar{q} ^2
e^{- 2\lambda |z|} \right| , \nonumber \\
& y = |z| - \ln \left(1 + \tilde{g}^2  e^{2|z|} \right),
\end{align}
with $\bar{q} = \tilde{q} /\sqrt{2}\lambda$. Now inserting metric functions
\eqref{setup-eq4} into the $zz-$component of Einstein equation, one finds the constraint
\begin{equation}\label{setup-eq85}
\lambda^2 +2 (\lambda_3+\lambda_4)^2 + \frac{16}{3} \lambda_3^2 = 1
\end{equation}
that reduces number of independent constants by one.

Introducing boundaries and including absolute value in the solution suggests some branes as delta function singularities which arise as second derivative of absolute values. A suitable configuration
of branes in the closed interval $[0,L]$ is \cite{Afshar:2009ps}:
\begin{align} \label{setup-eq9}
 T_{MN}^{\text{brane}} & = - \left[(T_4 + \tilde{T}_3) g_{\mu \nu} \delta^{\mu} _M \delta ^{\nu}_N
 + T_4 g_{\theta \theta} \delta^{\theta} _M \delta ^{\theta}_N \right] l_z e^{-v} \delta (z) \nonumber \\
 & - \left[(T_{L4} + \tilde{T}_{L3}) g_{\mu \nu} \delta^{\mu} _M \delta ^{\nu}_N
 + \tilde{T}_{L0} g_{00} \delta^0 _M \delta ^0_N +
 T_{L4} g_{\theta \theta} \delta^{\theta} _M \delta ^{\theta}_N \right] l_z e^{-v} \delta (L-z)
\end{align}
where $T_p(T_{Lp})$ stands for tension of $p-$brane located at $z=0(z=L)$ and \textit{tilde} denotes density of tension. In this configuration, 4-branes are boundaries of the space and 3 and zero branes are smeared over 4-branes. Inclusion of 3 and zero branes is essential for matching the energy-momentum and the Einstein tensors. 
The Israel junction conditions then read\footnote{For technical details in various branes configurations consult with \cite{Afshar:2013bg}.}
\begin{align} \label{setup-eq10}
[a ^{\prime}(z) - w^{\prime}(z)]_{z=0^+} & = 0  \nonumber \\
 [b ^{\prime}(z) - a^{\prime}(z)]_{z=0^+} & = - \kappa ^2 e^{v(0)} \tilde{T}_3  \nonumber \\
 [3 a ^{\prime}(z)+ w ^{\prime}(z)]_{z=0^+} & = - \kappa ^2 e^{v(0)} T_4  \nonumber \\
 [a ^{\prime}(z)- w ^{\prime}(z)]_{z=L^-} & = \kappa ^2 e^{v(L)} \tilde{T}_{L0}  \nonumber \\
 [b ^{\prime}(z)- a ^{\prime}(z)]_{z=L^-} & = \kappa ^2 e^{v(L)} \tilde{T}_{L3}  \nonumber \\
 [3 a ^{\prime}(z)+ w ^{\prime}(z)]_{z=L^-} & = \kappa ^2 e^{v(L)} T_{L4},
\end{align}
from which brane tensions can be derived and $[f(z)]_{z_0}$ is defined as
\begin{equation} \label{bracket-def}
[f(z)]_{z_0}=\lim_{\varepsilon\to 0} \left(f(z_0 +\varepsilon)-f(z_0-\varepsilon)\right).
\end{equation}
Since metric functions are even function of $z$ and we are working in the interval $[0,L]$, then on the boundary $z=0^+ (z=L^-)$ one should replace, for example, $a'(-\varepsilon)$ with $-a'(\varepsilon)$ while on the boundary $z=L^-$, $a'(L^+)$ should be replaced with $-a'(L^-)$. The first condition of Eqs.~(\ref{setup-eq10}) gives
\begin{eqnarray}
14 \lambda_3 + 6 \lambda_4 = -3 \alpha \lambda
\end{eqnarray}
where $\alpha=\frac{1 + \kappa^2 \bar{q}^2}{1 - \kappa^2 \bar{q}^2}$.
Using \eqref{setup-eq85}, we then get the following relations between $(\lambda_3, \lambda_4)$ and $(\lambda, q)$:
\begin{align} \label{setup-eq8}
\lambda_3^{\pm} = \frac{3}{20} \alpha \lambda \pm \frac{3}{40} \sqrt{20 - 20 \lambda^2 -6 \alpha^2 \lambda ^2}, \nonumber\\
\lambda_4^{\pm} = \frac{3}{20} \alpha \lambda \mp \frac{7}{40}\sqrt{20 - 20 \lambda ^2 -6 \alpha^2 \lambda ^2}.
\end{align}
The only remaining
constant to be noted is $z_3$ in \eqref{setup-eq4} that is essentially unimportant
and can be absorbed by rescaling coordinates. However, we keep this
constant for further simplification.

\section{The Null Energy Condition }\label{nullenergy}

\noindent Before study the gravitational perturbation in the above background, it is worth to pause for a while and consider the null energy condition.
This condition appears as a constraint for a matter distribution to be physical in the context of classical general relativity. It simply states that for any null vector $\xi^M$, the following inequality holds for the energy momentum tensor,
\begin{eqnarray}\label{null-T}
T_{MN}\xi^M\xi^N \geq 0.
\end{eqnarray}
Since $\xi$ is a null vector using the Einstein equation one finds,
\begin{eqnarray}\label{null-R}
R_{MN}\xi^M\xi^N \geq 0.
\end{eqnarray}
To be specific, let us choose $\xi^M=(e^{-w}, e^{-a}, 0, 0, 0, 0)$, so \eqref{null-R} turns to $-R^0_0+R^1_1\geq 0$. This condition is satisfied in the bulk as in the following \cite{Afshar:2009ps},
\begin{eqnarray}\label{null-eq3}
e^{2v}(-R^0_0+R^1_1)&=&w''-a''+(w'-a')^2+(b'-v')(w'-a')+4a'(w'-a') \geq 0 \nonumber\\
&=& w''-a''= x'' \geq 0
\end{eqnarray}
where the gauge condition is used. It is easy to verify that the last inequality $x''\geq 0$ is true.

However, this is not the whole story, since our model includes branes as boundaries. To investigate the null energy condition at boundaries, we apply it directly to branes energy-momentum tensor \eqref{setup-eq9},
\begin{eqnarray}
-T_{0}^{0} + T_1^1 =
+ \tilde{T}_{L0}l_z e^{-v} \delta (L-z) \geq 0
\end{eqnarray}
This condition implies $\tilde{T}_{L0} \geq 0$.
For further constraint, consider the null vector in \eqref{null-T} to be $\xi^M=(e^{-w}, 0, 0, 0, 0, e^{-b})$, then
\begin{eqnarray}
-T_{0}^{0} + T_\theta^\theta  =  \tilde{T}_{3}l_z e^{-v} \delta (z)
+ (\tilde{T}_{L3}+\tilde{T}_{L0})l_z e^{-v} \delta (L-z) \geq 0
\end{eqnarray}
It gives $\tilde{T}_{3}\geq 0$ and $\tilde{T}_{L3}+\tilde{T}_{L0} \geq 0$.

Thus in any physical solution to satisfy the null energy condition, $\tilde{T}_{3}$ should be negative or zero, while $\tilde{T}_{L0}$ and $\tilde{T}_{L3}+\tilde{T}_{L0}$ should be non-negative. To translate these conditions into some constraints on independent constants in the model, we firstly set $\tilde{T}_{L0}\ge 0$ in the fourth equation of Eqs.~(\ref{setup-eq10}). This, after a bit of algebra, gives $e^{-2 \lambda L} \le 1$ that is always true. 
Two conditions $\tilde{T}_3 \ge 0$ and $\tilde{T}_{L0}+\tilde{T}_{L3} \ge 0$ simplify commonly to the following inequality
\begin{align}
\pm \sqrt{20-20 \lambda ^2 -6 \alpha ^2 \lambda ^2} \le 3 \alpha \lambda ,
\end{align}
in which plus/minus signs originate from definitions of $\lambda^{\pm}_3$ and $\lambda^{\pm}_4$ in Eq.~(\ref{setup-eq8}). This inequality is satisfied unconditionally if $(\lambda_3,\lambda_4)=(\lambda_3^{-},\lambda_4^{-})$ while the choice $(\lambda_3,\lambda_4)=(\lambda_3^{+},\lambda_4^{+})$ gives rise to the constraint $\lambda ^2 \ge 4/(4+3\alpha^2)$. We therefore adopt the choice $(\lambda_3,\lambda_4)=(\lambda_3^{-},\lambda_4^{-})$ that is less sever.
%
%
\section{Small space-time fluctuations } \label{small}
\noindent To understand behaviour of graviton in this space-time, we consider small
fluctuations around the background metric.  Recalling the Palatini identity,
the small fluctuation $\delta g_{MN}$ implies a variation in Ricci tensor
as, to leading order in $\delta$,
\begin{equation} \label{small-eq1}
- \delta \mathcal{R}_{MN} = \frac{1}{2} g^{AB} \nabla _A (\nabla_M \delta g_{NB}
+\nabla_N \delta g_{MB}-\nabla_B \delta g_{MN})- \frac{1}{2} \nabla_M \nabla_N \delta g_A^A.
\end{equation}
We take the tensorial fluctuations in the spatial sector on brane, i.e.
$\delta g_{MN} = \delta g_{ij} \delta ^i_M \delta ^j_N$, and also adopt
the conventional transverse-traceless gauge in which $\delta g_i^i=0$ and
$\partial_i \delta g_k^i=0$. The immediate consequence of this gauge is that the
last term in \eqref{small-eq1} vanishes identically. The other terms simply
show that just the components $\delta \mathcal{R}_{zi} $ and $\delta \mathcal{R}_{ij}$
may be non-zero. Keeping in mind that the background metric depends only on $z$
coordinate, $\delta \mathcal{R}_{zi} $ is obtained to be zero as well.
The only remaining possibility is therefore
\begin{align} \label{small-eq2}
- \delta \mathcal{R}_{ij} & = -\frac{1}{2} \Box \delta g_{ij} +
\frac{1}{2} g^{kl} \nabla _k (\nabla_i \delta g_{jl}+\nabla_j \delta g_{il}) \nonumber \\
& = -\frac{1}{2} \Box \delta g_{ij} + a^{\prime} g^{zz} (\partial_z \delta g_{ij} -2 a^{\prime} \delta g_{ij})
\end{align}
where $\Box \equiv g^{AB} \nabla_A \nabla_B$ stands for the d'Alembert operator. We now
consider the right-hand side of the Einstein equation in \eqref{setup-eq2} as
\begin{equation}
-\delta \mathcal{R}_{MN}=\delta S_{MN}+\delta S^{\textmd{brane}}_{MN}
\end{equation}
where
\begin{equation}
S_{MN}=T_{MN}-\frac{T}{D-2}g_{MN},
\end{equation}
and then change the metric tensor as $g_{MN} \to g_{MN} + \delta g_{ij} \delta ^i_M \delta ^j_N$ to give
\begin{align} \label{small-eq3}
 \delta S_{MN} & = (\frac{g^2}{\kappa ^2} e^{\phi} - \frac{1}{12} \kappa^2 H^2 e^{- 2 \phi}) \delta g_{MN} \nonumber \\
& = - a^{\prime \prime} e^{-2v} \delta g_{MN}.
\end{align}
in which we have used the $(ii)-$component of Einstein field equations.
The only remaining contribution to the energy-momentum tensor to be taken into account
is that of branes. Recalling Einstein equation in \eqref{setup-eq2} and
energy-momentum tensor on the branes \eqref{setup-eq9}, then $\delta S_{MN}^{\textmd{brane}}$ is obtained as
\begin{equation} \label{small-eq4}
 \delta S_{MN}^{\textmd{brane}} = \kappa^2 ( \delta T_{MN} - \frac{\delta T}{D-2} g_{MN} - \frac{T}{D-2} \delta g_{MN}) .
\end{equation}
Notice that $T_{MN} \propto G_{MN}$ which leads to
$\delta T_{MN} \propto \delta g_{MN} =\delta g_{ij} \delta ^i_M \delta ^j_N $, and
$\delta T = \delta (G_{MN} T^{MN} ) \propto \delta g^M_M=0$ because of the tracless gauge.
Two other terms can be derived as follows,
\ba
\frac{T}{D-2}&=& -\tfrac{1}{D-2} \left(\left[4(T_4+\tilde{T}_3)+T_4 \right]e^{-v}\delta(z)  +\left[4(T_{L4}+\tilde{T}_{L3})+T_{L4}+\tilde{T}_{L0} \right]e^{-v}\delta(L-z)\right)  \nonumber\\
\delta T_{MN}&=& - (T_4+\tilde{T}_3)e^{-v}\delta g_{MN}\delta(z)-(T_{L4}+\tilde{T}_{L3})e^{-v}\delta g_{MN}\delta(L-z).
\ea
The branes contribution finally becomes
\begin{align} \label{small-eq5}
  \delta S_{MN}^{\textmd{brane}} & = \frac{\kappa ^2}{4} \left[ T_4 \delta(z)+(\tilde{T}_{0}
  + T_4) \delta (L-z)\right] e^{-v} \delta g_{MN} \nonumber\\
 & = - \frac{1}{2} \left[ (3 a^{\prime} + w^{\prime})(0^+) \delta(z)-
 4 a^{\prime}(L^-) \delta(L-z)\right] e^{-2v} \delta g_{MN} \nonumber\\
  & = - 2 \left[  a'(0^+) \delta(z)-
  a'(L^-) \delta(L-z)\right] e^{-2v} \delta g_{MN}
\end{align}
where we used Eqs.~\eqref{setup-eq10} and definition~\eqref{bracket-def}.
We now gather Eqns. \eqref{small-eq2}, \eqref{small-eq3} and \eqref{small-eq5} to get the equation governing fluctuations:
\begin{equation} \label{small-eq6}
 g_{zz} \Box \delta g_{ij} -2 a^{\prime} \partial_z \delta g_{ij} + (4 {a^{\prime}} ^2 - 2 a^{\prime \prime}) \delta g_{ij}
= 4 [ a^{\prime}(0^+) \delta(z)- a^{\prime}(L^-) \delta(L-z)] \delta g_{ij}.
\end{equation}
Notice that the function $a^{\prime \prime}$ here should be written as $a^{\prime \prime} \textrm{sign}(z)
+ 2a^{\prime} \delta (z)$ because of absolute value in its argument.
Since we have previously chosen the gauge $\partial_z (g^{zz} \sqrt{-g})=0$ in fixing coordinate $z$,
the d'Alembertian operator reduces to $ g^{MN} \partial_M \partial_N $ which simplifies \eqref{small-eq6}.
To recast this equation in the form of a Schr\"odinger-like one, we perform the transformation
$\delta \tilde{g}_{ij} = \delta g_{ij} e^{-a} $ and take the Fourier decomposition of the form
$\delta\tilde{g}_{ij} = \textrm{exp}(i \eta_{\mu \nu} \tilde{p}^{\mu} x^{\nu}) \psi(z) $ to get
\begin{align} \label{small-eq7}
\frac{d^2 \psi}{dz^2} +\left[3{a^{\prime}} ^2 -a^{\prime \prime} \textrm{sign}(z)
+e^{2v -2w} (E^2 -c^2 p^2)\right] \psi \nonumber \\
=6 [ a'(0^+) \delta(z)- a'(L^+) \delta(L-z)] \psi.
\end{align}
where we used $g_{zz} = l_z^2 e^{2v}$ and defined dimensionless energy $E:=\tilde{E} l_z$ and momentum $p:=\tilde{p} l_z$. Also, $c(z):=e^{w-a}$ that is, in terms of metric functions \eqref{setup-eq4},
\begin{equation} \label{small-eq8}
c(z)^2= \frac{e^{\lambda[\alpha z_3+(\alpha-1)|z|]}}{1- \bar{q}^2 e^{-2\lambda|z|}}.
\end{equation}
We now fix $z_3$ such that $c(0)^2=1$ implying $e^{\alpha \lambda z_3} = 1- \bar{q}^2$.
This choice also imposes a restriction on $ \bar{q}^2$ to be smaller than unity for which $c^2(z)>0$.
Then we rewrite $c^2(z)$ as,
\begin{equation} \label{small-eq8a}
c(z)^2= \frac{(1- \bar{q}^2)e^{(\alpha-1)\lambda|z|}}{1- \bar{q}^2 e^{-2\lambda|z|}}.
\end{equation}
In eq. \eqref{small-eq7}, a factor $e^{2(\lambda_3 + \lambda_4)z_3}$ is included in the function $e^{2v-2w}$ that
can be absorbed in $E$ and $p$ by rescaling. \\
To find boundary conditions, we integrate Eqn. \eqref{small-eq7} over a small neighbor around
boundaries at $z=0$ and $z=L$. The resulting conditions are
\begin{align} \label{small-eq9}
\psi' (0^+)= 3a'(0^+) \psi(0), \nonumber \\
\psi' (L^-)= 3a'(L^-) \psi(L).
\end{align}
 Having found boundary conditions we now
proceed to find a solution in the bulk. However, the complication in the potential of \eqref{small-eq7} leads us to numerical methods.\\
Before restricting ourselves to any special values of constants, it is worth to make sense
of dispersion relation by rewrite Eq.~\eqref{small-eq7} in the bulk as
 \begin{equation} \label{small-eq10}
          - \psi^{\prime \prime} + \hat{q}(z)\psi = \hat{\lambda} \hat{w}(z) \psi,
  \end{equation}
  where we have defined the eigenvalue $\hat{\lambda}=E^2$, weight function $\hat{w}(z)=e^{2v-2w} > 0$,
  and $\hat{q}(z)= p^2 c^2 \hat{w}+a^{\prime\prime}-3{a^{\prime}}^2$. The weight function suggests
  to adopt the normalization of wave function as $\int_{0}^{L} \psi^{\star} \hat{w} \psi dz=1 $,
  and consequently define the expectation value of a given function $f(z)$ as
  $\left <f \right >:= \int_{0}^{L} \hat{w} \psi^{\star} f  \psi dz $.
We now multiply Eq.~\eqref{small-eq10} by $\psi^{\star}$, complex conjugate of wave function,
 and integrate the result from $z=0$ to $L$ to obtain the well-known Green's first identity
\begin{equation} \label{small-eq11}
  \hat{\lambda} \int_{0}^{L} \hat{w}\psi \psi^{\star} \,dz = \int_{0}^{L} \psi^{\star} (\mathcal{L}\psi)\,dz
  = \psi^{\prime} \psi^{\star} |_{L}^{0}+ \int_{0}^{L}(\psi^{\prime} \psi^{\star \prime}+\hat{q} \psi \psi^{\star})\,dz.
\end{equation}
Inserting corresponding quantities and functions in this identity, one finds
\begin{equation} \label{small-eq12}
      E^2 = p^2 \left<c^2 \right> + \left < \hat{w}^{-1} (a^{\prime \prime}-3{a^{\prime}}^2) \right>
      + \psi^{\prime} \psi^{\star} |_{L}^{0}+ \int_{0}^{L} \psi^{\prime} \psi^{\star \prime}\,dz.
\end{equation}
This is an energy-momentum dispersion relation for which the group velocity $v_g = dE/dp$ times the phase velocity $v_{ph}=E/p$ reads as,
 \begin{equation} \label{small-eq13}
       \frac{E}{p} \frac{dE}{dp} = \left<c^2\right>.
 \end{equation}
 On the brane localized at $z=0$ we have $c(0)=1$, so this equality
 reduces to the familiar relation $v_{ph} v_{g}=1$.
 However, $c$ in the r.h.s of this equality is no longer constant in the bulk which
 leads to a superluminal behaviour of graviton in this model.

 Although  $\left<c^2\right>$ cannot be determined unless we have the
 exact form of wave function in hand, it is possible to estimate upper and lower bounds to this quantity. The expectation value $\left< c^2 \right>$
 is in fact weighted average of function $c^2 (z) $ with the (\textit{normalized positive})
 measure $ \hat{w} {|\psi|}^2$, probability density function. Based on the fact that
 average of any function over an interval lies between its extrema in that interval,
 we can write $\textit{min} \{c^2(z) \} \leq \left <c^2\right > \leq \textit{max} \{c^2(z) \} $
 for $z\in [0,L]$. The equal sign occurs when the function $c(z)$ is constant over
 the interval that is not the case we are considering. We then can write
\begin{equation} \label{small-eq14}
 \min \{ c^2 \} \le \frac{E}{p} \frac{dE}{dp} \le \max \{ c^2 \}.
 \end{equation}
To find extrema of $c^2(z)$ defined by \eqref{small-eq8a}, we notice that this function is
strictly increasing meaning that its derivative is positive for all $z$ in the domain
$0\le z\le L$. This observation ensures us that the extrema occur at endpoints $z=0$ or $z=L$. We therefore can safely write $\textit{min} \{c^2\} = c^2(0)=1$ and $\textit{max} \{c^2\} = c^2(L)$.
Inserting these values in the inequality \eqref{small-eq14}, it becomes
 \begin{equation} \label{small-eq15}
 1 \le \frac{E}{p} \frac{dE}{dp} \le \left (\frac{1- \bar{q}^2}{1-\bar{q}^2 e^{-2\lambda L}}\right )
 e^{(\alpha-1)\lambda L},
  \end{equation}
 The constants $\bar{q}$ and $\lambda$ here refer to contributions of electric $H$ field and dilation
 to the dispersion relation while the effect of cosmological constant does not appear explicitly.
 This relation determines the most speed violation from speed of light for a given set of constants.
In particular, the problem becomes non-dispersive if either $\bar{q}=0$ or $\lambda =0$,
 and the r.h.s approaches to infinity for large $L$ limit.

We can now solve the equation \eqref{small-eq7} for dimensionless quantities $(z,E,p)$ and thereafter interpret them as $(\eta/l_z, \tilde{E} l_z,\tilde{p} l_z)$. To find 6D Planck mass, we integrate over extra dimensions of the action \eqref{setup-eq1} as
\begin{align}\label{6D-Planck-mass}
S_6 &= M^4_{(6)} \int \sqrt{-G}R^{(6)} d^6 x \nonumber \\
& = M^4_{(6)} \int d^4 x \sqrt{-g} \left(-R_{00}^{(4)} \int d\theta' d\eta \sqrt{G} e^{-2w} +\delta^{ik} R_{ik}^{(4)} \int d\theta' d\eta \sqrt{G} e^{-2a}\right) 
\end{align}
where $M^4_{(6)}=\tfrac{1}{2\kappa^2_6}$, $g$ and $G$ are respectively determinants of 4D flat metric and 6D metric.
Since $e^{-2w} = c^2 e^{-2a}$, two integrals in r.h.s. are approximately equal for sufficiently small violation of speed from unity, say $\varepsilon:=c-1$. In this regime, one can define $V_2:=\int d\theta' d\eta \sqrt{G} e^{-2w}$ as the volume of 2D compactified space and obtain
\begin{align} 
S_6 = M^4_{(6)} V_2\int \sqrt{-g}R^{(4)} d^4 x \equiv  M^2_{(4)} \int \sqrt{-g}R^{(4)} d^4 x
\end{align}
with $M^2_{(4)}:=\tfrac{1}{2\kappa^2_{4}}$. As a result of this relation, the 6D Planck mass is obtained as $M^4_{(6)}= M^2_{(4)}/V_2$ with $M_{(4)} = 2\times 10^{18} \text{GeV}$. We use this relation in the next section.

\section{Numerical results} \label{numer}
\noindent To solve equation \eqref{small-eq7} numerically, we firstly study the constraints
on constants involved. The charge $q$ and coupling constant $g$
seem to be arbitrary everywhere, as expected from a physical point of view.
Returning to metric functions \eqref{setup-eq4}, one finds that a real metric tensor implies
that  both of $\lambda_3$ and $\lambda_4$ in \eqref{setup-eq8} are real.
This condition imposes a constraint on $\lambda$ as
\begin{equation} \label{upper bound}
      \lambda \le \sqrt{\left(1+\tfrac{3}{10} \alpha^2\right)^{-1}}.
\end{equation}
The final constant to be specified is the distance separating the branes, $L$. Notice that $L$ is not fixed in this model. Instead it is chosen phenomenologically to fit experimental bounds as explained below. It is possible to study this radial mode and its spectrum as well. Since the radion field propagates in the bulk, we expect that its massless mode, if any, violates the Lorentz symmetry. However in this article we focus on the tensorial perturbation and postpone the radial one for future works. 

Once the set of constants ($\lambda, \tilde{q}, \tilde{g}, L$) is fixed, for every value of momentum $p$, boundary conditions are satisfied just for some special values of energy $E$. Then the mass spectrum of graviton can be obtained by finding energies correspond to zero momentum limit. Especially, the massless graviton is of great interest and it does exist provided the \textit{smallest} energy approaches
to zero when momentum does so. This statement may be considered as a criteria for fixing either $L$ or $\lambda$, given other constants.

Among all possible configurations, we are interested in the case that all tensions are non-negative. As said before, the choice $(\lambda_3 , \lambda_4) = (\lambda_3^- , \lambda_4^-)$ guarantees that $T_{L0}$, $T_{L3}$ and $T_3$ are non-negative and consequently null energy conditions are satisfied. Furthermore, the inequality $T_4 \ge 0$ reads the following condition
\begin{equation}  \label{lowe bound g}
1-\tfrac{4}{5} \alpha \lambda - \tfrac{1}{10} \sqrt{20 - 20 \lambda ^2 - 6 \alpha ^2 \lambda ^2} \ge \frac{2}{1 + \tilde{g}^2}
\end{equation}
which imposes a lower bound on $\tilde{g}^2$, provided the left-hand side itself is non-negative that is so if
\begin{equation}  \label{lower bound}
\lambda \le \frac{8\alpha - \sqrt{8(\alpha ^2 -2)}}{2 + 7 \alpha ^2}.
\end{equation}
This inequality now implies that $\alpha \ge \sqrt{2}$ or $\bar{q}^2 \ge 0.17$.
We now have two conditions \eqref{upper bound} and \eqref{lower bound} on $\lambda$ that reduces to \eqref{lower bound}. Therefore, the constant $\lambda$ can be written as
\begin{equation}  \label{lambda}
\lambda = \mu \left(\frac{8 \alpha -\sqrt{8(\alpha ^2 -2)}}{2 + 7 \alpha ^2} \right)
\end{equation}
for $0<\mu <1$ being a fine-tuned parameter satisfying the criteria above.
Finally $T_{L4}$ can be checked easily to be non-negative where $T_4$ does so.
In this manner we firstly fix $\bar{q}^2$ and put $\tilde{g}^2$ twice of that obtained from equality sign of \eqref{lowe bound g} and then search for suitable $\mu$ in \eqref{lambda}. This strategy leaves $L$ unconstrained
 and the violation of speed from unity, $\varepsilon$, may take every value due to inequality \eqref{small-eq15}.
 
 However, there have been reported some constraint on the size of violation of
 graviton's propagation speed by general relativity tests in solar system and binary pulsar \cite{Will:2001mx}
 that is about $\varepsilon \le 10^{-6}$.
 Recalling equation \eqref{small-eq15}, this upper bound of $\varepsilon$
 is translated as a constraint on $\lambda L$.
 It is easy to check that
 for small $\lambda L \ll 1$, this equation reads $c^2(\lambda L) = 1+O(\lambda^2 L^2)+...$,
 in which the ellipses indicates higher orders of $\lambda L$. Here we change both $\lambda$ and $L$ under the criteria that massless graviton does exist and the upper bound $\lambda L \approx 10^{-3}$ that gives $\varepsilon \approx 10^{-6}$.

\begin{figure}[ht]
\centering
\includegraphics[scale=0.5]{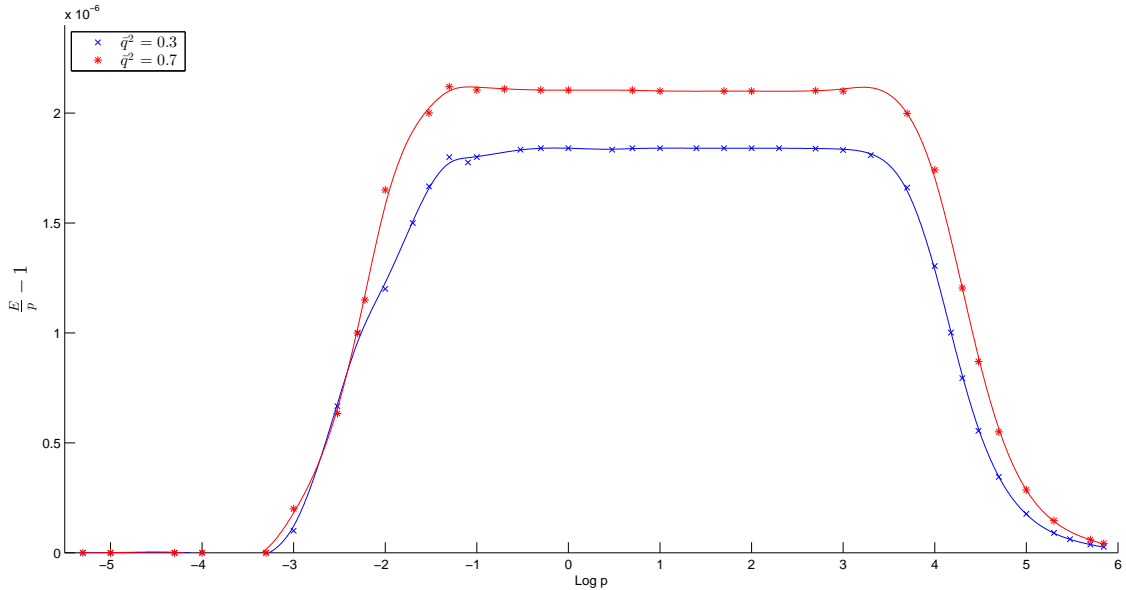}
\caption{\footnotesize Speed of graviton versus momentum in log scale. Cross/asterisk symbols denote chosen momenta. The curve indicates that in a wide range of momenta, the graviton violates the speed limit.}
\label{figure1}
\end{figure}

Two set of constants obtained in this way are $(\bar{q}^2, \tilde{g}^2, L, \lambda)=(0.3, 5.322, 1.258, 4.38\times 10^{-3})$
and $(\bar{q}^2, \tilde{g}^2, L, \lambda)=(0.7, 5.315, 1.257, 1.32\times 10^{-3})$.
We will refer to each set of constants by its $\bar{q}^2$-value.
Inserting these values, we chose momentum in the interval $[0, 7 \times 10^5]$ and changed energy,
by the increment $\delta E$, from zero to the value satisfying boundary conditions.
To be more accurate, the energy increment was chosen in two regimes:
$\delta E=10^{-8}$ for $p\in [0,1]$, $\delta E=10^{-6}$ for remaining part of interval.
Since momentum varies in a wide range of, we used logarithmic scale for momentum. For each set of constants, the violation from speed of light ($\tfrac{E}{p}-1$) is shown in figure (\ref{figure1}).
%
%

As we can see, the graviton has a similar behaviour in both examples:
It begins with the speed of light for small momenta and then its velocity increases to a maximum.
For a wide range of momenta, the velocity remains nearly constant at this maximum and thereafter falls off to unity asymptotically.
The maximum value of $c^2(z)$ in the r.h.s of inequality \eqref{small-eq15} is obtained
$c^2(L)-1=1.847\times 10^{-5}$ for the first set of constants ($\bar{q}^2=0.3$), and $c^2(L)-1=2.112\times 10^{-5}$ for the other set.
For a massless particle ($v_{ph} =v_g$), this inequality implies that
$v_{g}|_{\bar{q}^2=0.3}-1<1.847\times 10^{-5}$ and $v_{g}|_{\bar{q}^2=0.7}<2.112\times 10^{-5}$ that is verified by figure (\ref{figure1}).
The tension of each brane is also shown in the Table (\ref{table-tension}) which ensures us that null energy conditions are satisfied.
\begin{table}[h]
\small
\begin{center}
  \begin{tabular} { | c | c | c | c | c | c | c }
    \hline
    Tension($\times \kappa^{-2} e^{\lambda_3 z_3} $)  &  $\bar{T}_3$  & $T_4$ & $T_{L0}$ & $\bar{T}_{L3}$ & $T_{L4}$  \\
    \hline
    $\bar{q}^2=0.3$   & 19.712  &  5.039  &  0.004  &  118.526  & 68.065 \\
    \hline
    $\bar{q}^2=0.7$   & 24.317  & 6.225  &  0.005  &  146.080 & 84.003  \\
        \hline
  \end{tabular}
    \caption{\footnotesize Tension of branes for each set of constants}
    \label{table-tension}
\end{center}
\end{table}

\noindent Similar to the case of Randall-Sundrum model, the positive $T_{4}$ guarantees that Newtonian gravity can be recovered on the $4-$brane located at $z=0$.
It is also worth to find the mass gap between zero mode and some lowest massive modes that are listed in the following table:
\begin{table}[h]
\small
\begin{center}
  \begin{tabular} { | c | c | c | c | c | c | c }
    \hline
    Mass spectrum &  $E_0$  & $E_1$ & $E_2$ & $E_3$ & $E_4$   \\
    \hline
    $\bar{q}^2=0.3$   & 0  &  42.842  &  79.061  & 115.802  & 152.879 \\
    \hline
    $\bar{q}^2=0.7$   & 0  & 65.391  &  120.653  &  176.725 & 233.307  \\
        \hline
  \end{tabular}
    \caption{\footnotesize Mass spectrum of graviton in dimensionless variable $E=\tilde{E} l_z$.} Taking $l_z\sim  \text{TeV}^{-1}$ turns the mass spectrum into TeV units. 
    \label{table-mass spectrum}
\end{center}
\end{table}

The appearance of a mass gap would be interesting phenomenologically. To make sense of order of magnitude of energy levels, we notice that the dimensionless factor $E$ here is in fact $\tilde{E} l_z$. Hence, the energies are of order $l_z^{-1}$. Taking $l_z\sim  \text{TeV}^{-1}$ turns the mass spectrum into TeV units, so phenomenologically consistent with observation bounds on massive gravitons. Notice that we have ignored the $\theta$-direction KK modes of graviton in equation \eqref{small-eq7}, so we expect $l_\theta$ to be much smaller than $l_z$. 

As the last quantity we consider 6D Planck mass. The graviton contribution to the Loop corrections to standard model particles gives a bound on the graviton dispersion relation \cite{Burgess:2002tb}. This loop correction bound depends on $M_{(6)}$ and would be stronger than $\varepsilon \le 10^{-6}$ by the solar system observation, only if $M_{(6)}$ is not far above TeV scale. 

Calculation of the 6D Planck mass gives  
$M_{(6)} = \Gamma(\bar{q},\tilde{g},L,\lambda)\sqrt{l_z^{-1} l_{\theta}^{-1}}$, with $\Gamma(\bar{q},\tilde{g},L,\lambda)$ comes from the volume of compactified 2D space, $V_2$, in~\eqref{6D-Planck-mass} that is $\Gamma(\bar{q}^2 = 0.3) = 1.765 \times 10^{-3}$ and $\Gamma(\bar{q}^2 = 0.7) = 1.675 \times 10^{-3}$. Now assuming $\sqrt{l_z^{-1} l_{\theta}^{-1}} \sim  10^{4} \text{TeV}^{-1}$, 
we get to $M_{(6)} \sim 10 \text{TeV}$ that is in the order of magnitude not to impose stronger graviton loop correction bound than $\varepsilon \le 10^{-6}$ \cite{Burgess:2002tb}. 
%
%
\section{Conclusion}
We have considered the dispersion relation for gravitational wave in the six-dimensional space
compactified to 4D, in the presence of dilaton and an electric $H$ field.
The dispersion relation seems to depend on the charge and the dilaton coupling constant as
well as an additional integration constant to be fine-tuned in the model. We have determined the constant
under the condition that the model contains a massless graviton, beside massive modes which are high enough to satisfy experimental bounds. The compactified lengths order of magnitude were chosen such that the graviton dispersion relation to be consistent with direct observations bounds as well as its contribution to the standard model particle propagator loop corrections. Any radial perturbation of the background which may fix the separation of two branes and presumably show a Lorentz violating behavior is left for future studies.  

We take two numerical examples and found that
the graviton moves at speed of unity for small momenta. As the momentum increases the speed
experiences a rapid change and get to a maximum greater than unity, the speed of light.
For a large interval of momenta, the speed remains approximately constant at the maximum,
and finally it approaches to unity asymptotically. On the other hand, since standard model fields are confined on the brane at $z=0$ where $c=1$, they don't expertise any Lorentz violating dispersion relation.

This model provides an example of asymmetric time warp compactification which presents Lorentz violation for gravitational waves while the standard model fields well behaved with Lorentz symmetry. This is achieved despite of a no-go theorem according to which in $D \neq 6$, no compactification with asymmetric time warping exists unless violates the null energy condition. Hereby we presented a model in which the null energy condition is satisfied and the speed limit is exceeded 1 for gravitational waves as a sign of gravitational Lorentz violation. This model can be an example (or candidate) for any situation where the Lorentz violation is interesting either theoretically or experimentally.


\acknowledgments{
Authors would like to thank Hamid Afshar and Hassan Firouzjahi for valuable discussions and reading the manuscript.}



\begin{thebibliography}{10}


\bibitem{ArkaniHamed:1998rs}
N.~Arkani-Hamed, S.~Dimopoulos, and G.~Dvali, ``{The Hierarchy problem and new
  dimensions at a millimeter},'' {\em Phys.Lett.} {\bf B429} (1998) 263--272,
\href{http://www.arXiv.org/abs/hep-ph/9803315}{{\tt hep-ph/9803315}}.

\bibitem{Randall:1999ee}
L.~Randall and R.~Sundrum, ``{A Large mass hierarchy from a small extra
  dimension},'' {\em Phys.Rev.Lett.} {\bf 83} (1999) 3370--3373,
\href{http://www.arXiv.org/abs/hep-ph/9905221}{{\tt hep-ph/9905221}}.

\bibitem{Randall:1999vf}
L.~Randall and R.~Sundrum, ``{An Alternative to compactification},'' {\em
  Phys.Rev.Lett.} {\bf 83} (1999) 4690--4693,
\href{http://www.arXiv.org/abs/hep-th/9906064}{{\tt hep-th/9906064}}.

\bibitem{ArkaniHamed:2000eg}
N.~Arkani-Hamed, S.~Dimopoulos, N.~Kaloper, and R.~Sundrum, ``{A Small
  cosmological constant from a large extra dimension},'' {\em Phys.Lett.} {\bf
  B480} (2000) 193--199,
\href{http://www.arXiv.org/abs/hep-th/0001197}{{\tt hep-th/0001197}}.

\bibitem{Kachru:2000hf}
S.~Kachru, M.~B. Schulz, and E.~Silverstein, ``{Selftuning flat domain walls in
  5-D gravity and string theory},'' {\em Phys.Rev.} {\bf D62} (2000) 045021,
\href{http://www.arXiv.org/abs/hep-th/0001206}{{\tt hep-th/0001206}}.

\bibitem{Nishino:1984gk}
  H.~Nishino and E.~Sezgin,
  ``Matter And Gauge Couplings Of N=2 Supergravity In Six-Dimensions,''
  Phys.\ Lett.\  B {\bf 144}, 187 (1984).


\bibitem{Salam:1984cj}
  A.~Salam and E.~Sezgin,
  ``Chiral Compactification On Minkowski $\times S^2$ Of N=2 Einstein-Maxwell
  Supergravity In Six-Dimensions,''
  Phys.\ Lett.\  B {\bf 147}, 47 (1984).


\bibitem{RandjbarDaemi:1985wc}
  S.~Randjbar-Daemi, A.~Salam, E.~Sezgin and J.~A.~Strathdee,
  ``An Anomaly Free Model In Six-Dimensions,''
  Phys.\ Lett.\  B {\bf 151}, 351 (1985).

\bibitem{Nishino:1986dc}
  H.~Nishino and E.~Sezgin,
  ``The Complete N=2, D = 6 Supergravity With Matter And Yang-Mills
  Couplings,''
  Nucl.\ Phys.\  B {\bf 278}, 353 (1986).

\bibitem{Carroll:2003db}
  S.~M.~Carroll and M.~M.~Guica,
  ``Sidestepping the cosmological constant with football-shaped extra
  dimensions,''
  arXiv:hep-th/0302067.

\bibitem{Navarro:2003vw}
  I.~Navarro,
  ``Codimension two compactifications and the cosmological constant  problem,''
  JCAP {\bf 0309}, 004 (2003)
  [arXiv:hep-th/0302129].

\bibitem{Navarro:2003bf}
  I.~Navarro,
  ``Spheres, deficit angles and the cosmological constant,''
  Class.\ Quant.\ Grav.\  {\bf 20}, 3603 (2003)
  [arXiv:hep-th/0305014].

\bibitem{Aghababaie:2003wz}
  Y.~Aghababaie, C.~P.~Burgess, S.~L.~Parameswaran and F.~Quevedo,
  ``Towards a naturally small cosmological constant from branes in 6D
  supergravity,''
  Nucl.\ Phys.\  B {\bf 680}, 389 (2004)
  [arXiv:hep-th/0304256].

\bibitem{Aghababaie:2003ar}
  Y.~Aghababaie {\it et al.},
  ``Warped brane worlds in six dimensional supergravity,''
  JHEP {\bf 0309}, 037 (2003)
  [arXiv:hep-th/0308064].

\bibitem{Gibbons:2003di}
  G.~W.~Gibbons, R.~Gueven and C.~N.~Pope,
  ``3-branes and uniqueness of the Salam-Sezgin vacuum,''
  Phys.\ Lett.\  B {\bf 595}, 498 (2004)
  [arXiv:hep-th/0307238].

\bibitem{Burgess:2004dh}
  C.~P.~Burgess, F.~Quevedo, G.~Tasinato and I.~Zavala,
  ``General axisymmetric solutions and self-tuning in 6D chiral gauged
  supergravity,''
  JHEP {\bf 0411}, 069 (2004)
  [arXiv:hep-th/0408109].

\bibitem{Cvetic:2003xr}
  M.~Cvetic, G.~W.~Gibbons and C.~N.~Pope,
  ``A string and M-theory origin for the Salam-Sezgin model,''
  Nucl.\ Phys.\  B {\bf 677}, 164 (2004)
  [arXiv:hep-th/0308026].


\bibitem{Lee:2005az}
  H.~M.~Lee and C.~Ludeling,
  ``The general warped solution with conical branes in six-dimensional
  supergravity,''
  JHEP {\bf 0601}, 062 (2006)
  [arXiv:hep-th/0510026].


\bibitem{Copeland:2007ur}
  E.~J.~Copeland and O.~Seto,
  ``Dynamical solutions of warped six dimensional supergravity,''
  JHEP {\bf 0708}, 001 (2007)
  [arXiv:0705.4169 [hep-th]].

\bibitem{Tolley:2006ht}
  A.~J.~Tolley, C.~P.~Burgess, C.~de Rham and D.~Hoover,
  ``Scaling solutions to 6D gauged chiral supergravity,''
  New J.\ Phys.\  {\bf 8}, 324 (2006)
  [arXiv:hep-th/0608083].

\bibitem{Burgess:2006ds}
  C.~P.~Burgess, C.~de Rham, D.~Hoover, D.~Mason and A.~J.~Tolley,
  ``Kicking the rugby ball: Perturbations of 6D gauged chiral supergravity,''
  JCAP {\bf 0702}, 009 (2007)
  [arXiv:hep-th/0610078].

\bibitem{Lee:2008pz}
  H.~M.~Lee,
  ``Flux compactifications and supersymmetry breaking in 6D gauged
  supergravity,''
  Mod.\ Phys.\ Lett.\  A {\bf 24}, 165 (2009)
  [arXiv:0812.3373 [hep-th]].

\bibitem{Parameswaran:2006db} 
  S.~L.~Parameswaran, S.~Randjbar-Daemi and A.~Salvio,
  ``Gauge Fields, Fermions and Mass Gaps in 6D Brane Worlds,''
  Nucl.\ Phys.\ B {\bf 767}, 54 (2007)
  [hep-th/0608074].

\bibitem{Parameswaran:2007cb} 
  S.~L.~Parameswaran, S.~Randjbar-Daemi and A.~Salvio,
  ``Stability and negative tensions in 6D brane worlds,''
  JHEP {\bf 0801}, 051 (2008)
  [arXiv:0706.1893 [hep-th]].

\bibitem{Parameswaran:2009bt} 
  S.~L.~Parameswaran, S.~Randjbar-Daemi and A.~Salvio,
  ``General Perturbations for Braneworld Compactifications and the Six Dimensional Case,''
  JHEP {\bf 0903}, 136 (2009)
  [arXiv:0902.0375 [hep-th]].

\bibitem{Salvio:2009mp} 
  A.~Salvio,
  ``Brane Gravitational Interactions from 6D Supergravity,''
  Phys.\ Lett.\ B {\bf 681}, 166 (2009)
  [arXiv:0909.0023 [hep-th]].

\bibitem{Afshar:2009ps}
H.~R. Afshar and S.~Parvizi, ``{3-Form Flux Compactification of Salam-Sezgin
  Supergravity},'' {\em JHEP} {\bf 0910} (2009) 086
[arXiv:0909.0023 [hep-th]].

\bibitem{Minamitsuji:2010fp}
  M.~Minamitsuji, N.~Ohta and K.~Uzawa,
  ``Dynamical solutions in the 3-Form Field Background in the Nishino-Salam-Sezgin Model,''
  Phys.\ Rev.\ D {\bf 81}, 126005 (2010)
  [arXiv:1003.5967 [hep-th]].

\bibitem{Minamitsuji:2010kb}
  M.~Minamitsuji, N.~Ohta and K.~Uzawa,
  ``Cosmological intersecting brane solutions,''
  Phys.\ Rev.\ D {\bf 82}, 086002 (2010)
  [arXiv:1007.1762 [hep-th]].

\bibitem{Kaelbermann:1998hu}
  G.~Kaelbermann and H.~Halevi,
  ``Nearness through an extra dimension,''
  gr-qc/9810083.

\bibitem{Visser:1985qm}
  M.~Visser,
  ``An Exotic Class of Kaluza-Klein Models,''
  Phys.\ Lett.\ B {\bf 159}, 22 (1985)
  [hep-th/9910093].

\bibitem{Chung:1999xg}
  D.~J.~H.~Chung and K.~Freese,
  ``Can geodesics in extra dimensions solve the cosmological horizon problem?,''
  Phys.\ Rev.\ D {\bf 62}, 063513 (2000)
  [hep-ph/9910235].

\bibitem{Kraus:1999it}
  P.~Kraus,
  ``Dynamics of anti-de Sitter domain walls,''
  JHEP {\bf 9912}, 011 (1999)
  [hep-th/9910149].

\bibitem{Youm:2001sw}
  D.~Youm,
  ``Brane world cosmologies with varying speed of light,''
  Phys.\ Rev.\ D {\bf 63}, 125011 (2001)
  [hep-th/0101228].

\bibitem{Kiritsis:1999tx}
  E.~Kiritsis,
  ``Supergravity, D-brane probes and thermal superYang-Mills: A Comparison,''
  JHEP {\bf 9910}, 010 (1999)
  [hep-th/9906206].

\bibitem{Alexander:1999cb}
  S.~H.~S.~Alexander,
  ``On the varying speed of light in a brane induced FRW universe,''
  JHEP {\bf 0011}, 017 (2000)
  [hep-th/9912037].

\bibitem{Bowcock:2000cq}
  P.~Bowcock, C.~Charmousis and R.~Gregory,
  ``General brane cosmologies and their global space-time structure,''
  Class.\ Quant.\ Grav.\  {\bf 17}, 4745 (2000)
  [hep-th/0007177].

\bibitem{Csaki:2000dm}
  C.~Csaki, J.~Erlich and C.~Grojean,
  ``Gravitational Lorentz violations and adjustment of the cosmological constant in asymmetrically warped space-times,''
  Nucl.\ Phys.\ B {\bf 604}, 312 (2001)
  [hep-th/0012143].

\bibitem{Cline:2001yt}
  J.~M.~Cline and H.~Firouzjahi,
  ``No go theorem for horizon-shielded self tuning singularities,''
  Phys.\ Rev.\ D {\bf 65}, 043501 (2002)
  [hep-th/0107198].

\bibitem{Cline:2003xy}
  J.~M.~Cline and L.~Valcarcel,
  ``Asymmetrically warped compactifications and gravitational Lorentz violation,''
  JHEP {\bf 0403}, 032 (2004)
  [hep-ph/0312245].

\bibitem{Dubovsky:2001fj}
  S.~L.~Dubovsky,
  ``Tunneling into extra dimension and high-energy violation of Lorentz invariance,''
  JHEP {\bf 0201}, 012 (2002)
  [hep-th/0103205].

\bibitem{Deffayet:2001aw}
  C.~Deffayet, G.~R.~Dvali, G.~Gabadadze and A.~Lue,
  ``Brane world flattening by a cosmological constant,''
  Phys.\ Rev.\ D {\bf 64}, 104002 (2001)
  [hep-th/0104201].

\bibitem{Frey:2003jq}
  A.~R.~Frey,
  ``String theoretic bounds on Lorentz violating warped compactification,''
  JHEP {\bf 0304}, 012 (2003)
  [hep-th/0301189].

\bibitem{Ganor:2006ub}
  O.~J.~Ganor,
  ``A New Lorentz violating nonlocal field theory from string-theory,''
  Phys.\ Rev.\ D {\bf 75}, 025002 (2007)
  [hep-th/0609107].

\bibitem{Koroteev:2007yp}
  P.~Koroteev and M.~Libanov,
  ``On Existence of Self-Tuning Solutions in Static Braneworlds without Singularities,''
  JHEP {\bf 0802}, 104 (2008)
  [arXiv:0712.1136 [hep-th]].

\bibitem{Gubser:2008gr}
  S.~S.~Gubser,
  ``Time warps,''
  JHEP {\bf 1001}, 020 (2010)
  [arXiv:0812.5107 [hep-th]].

\bibitem{Kiritsis:2012ta}
  E.~Kiritsis,
  ``Lorentz violation, Gravity, Dissipation and Holography,''
  JHEP {\bf 1301}, 030 (2013)
  [arXiv:1207.2325 [hep-th]].

\bibitem{Gubser:2011mp}
  S.~S.~Gubser,
  ``Superluminal neutrinos and extra dimensions: Constraints from the null energy condition,''
  Phys.\ Lett.\ B {\bf 705}, 279 (2011)
  [arXiv:1109.5687 [hep-th]].

\bibitem{Afshar:2013bg}
  H.~R.~Afshar, H.~Firouzjahi and S.~Parvizi,
  ``dS Solutions with co-dimension two branes in six dimensions,''
  Class.\ Quant.\ Grav.\  {\bf 30}, 235030 (2013)
  arXiv:1301.3160 [hep-th].
  
  \bibitem{Will:2001mx} 
    C.~M.~Will,
    ``The Confrontation between general relativity and experiment,''
    Living Rev.\ Rel.\  {\bf 4}, 4 (2001)
    [gr-qc/0103036].
  
  \bibitem{Burgess:2002tb} 
    C.~P.~Burgess, J.~M.~Cline, E.~Filotas, J.~Matias and G.~D.~Moore,
    ``Loop generated bounds on changes to the graviton dispersion relation,''
    JHEP {\bf 0203}, 043 (2002)
    [hep-ph/0201082].

\end{thebibliography}
\providecommand{\href}[2]{#2}\begingroup\raggedright\endgroup

\end{document}